\documentclass[prd,amsbook,showpacs,nofootinbib,tightenlines,preprintnumbers,superscriptaddress,twocolumn]{revtex4}
\usepackage{epsfig,graphics}
\usepackage{amssymb,amsmath,amsthm,graphicx,psfrag}

\newcommand{\beqn}{\begin{eqnarray}}
\newcommand{\eeqn}{\end{eqnarray}}
\newcommand{\be}{\begin{equation}}
\newcommand{\ee}{\end{equation}}

\begin{document}

\preprint{HU-EP-07/10}
\preprint{ITEP-LAT/2007-11}

\title{Calorons and dyons at the thermal phase transition \\
analyzed by overlap fermions}
\author{V.~G.~Bornyakov}
\affiliation{Institute for High Energy Physics,
Protvino, 142281, Russia}
\affiliation{Institute for Theoretical and Experimental Physics,
B. Cheremushkinskaya 25, Moscow 117259, Russia}
\author{E.-M.~Ilgenfritz}
\affiliation{Institut f\"ur Physik, Humboldt-Universit\"at zu Berlin,
Newtonstr. 15, D-12489 Berlin, Germany}
\author{B.~V.~Martemyanov}
\affiliation{Institute for Theoretical and Experimental Physics,
B. Cheremushkinskaya 25, Moscow 117259, Russia}
\author{S.~M.~Morozov}
\affiliation{Institute for Theoretical and Experimental Physics,
B. Cheremushkinskaya 25, Moscow 117259, Russia}
\author{M.~M\"uller-Preussker}
\affiliation{Institut f\"ur Physik, Humboldt-Universit\"at zu Berlin,
Newtonstr. 15, D-12489 Berlin, Germany}
\author{A.~I.~Veselov}
\affiliation{Institute for Theoretical and Experimental Physics,
B. Cheremushkinskaya 25, Moscow 117259, Russia}

\begin{abstract}
In a pilot study, we use the topological charge density defined by the 
eigenmodes of the overlap Dirac operator (with ultraviolet filtering by 
mode-truncation) to search for lumps of topological charge in $SU(2)$ 
pure gauge theory. Augmenting this search with periodic and antiperiodic 
temporal boundary conditions for the overlap fermions, we demonstrate that 
the lumps can be classified either as calorons or as separate caloron constituents 
(dyons). Inside the topological charge clusters the (smeared) Polyakov loop 
is found to show the typical profile characteristic for calorons and dyons. 
This investigation, motivated by recent caloron/dyon model studies, is performed 
at the deconfinement phase transition for $SU(2)$ gluodynamics on $20^3\times6$ 
lattices described by the tadpole improved L\"uscher-Weisz action. The transition
point has been carefully located. As a necessary condition for the caloron/dyon 
detection capability, we check that the LW action, in contrast to the Wilson 
action, generates lattice ensembles, for which the overlap Dirac eigenvalue 
spectrum smoothly behaves under smearing and under the change of the boundary 
conditions. 
\end{abstract}


\maketitle


\section{Introduction}
\label{sec:introduction}

The two current confinement scenarios, the 
mo\-no\-po\-le~\cite{Hooft1976,Mandelstam:1974pi,'tHooft:1981ht}
and the vortex mechanism~\cite{'tHooft:1977hy,Mack:1980rc}
of confinement in $SU(N)$ gauge theory have become unified 
within the $Z(N)$ vortex 
picture~\cite{'tHooft:2003ih,Greensite:2003bk,'tHooft:2004th, DiGiacomo:2005yq,Alkofer:2006fu,Boyko:2006ic}. 
Yet there exists the old hope to connect confinement also 
with the topological structure as understood in terms of 
instantons~\cite{Fukushima:1996yn,Fukushima:1997rc,Fukushima:2000ix},
calorons~\cite{Gerhold:2006sk}, dyons~\cite{Diakonov:2002fq,Diakonov:2007nv}, 
merons~\cite{Lenz:2003jp} and more generic objects~\cite{Wagner:2006qn},
all being carriers of Pontryagin charge.~\footnote{For brevity, in this
paper we understand ``calorons'', ``dyons'' and ``selfdual'' as including
also ``anticalorons'', ``antidyon'' and ``antiselfdual''.}
The main reason for this desire is to bring confinement, on first sight a rather 
abstract property of pure (lattice) Yang-Mills theory, in closer relation to the 
physical origin of chiral symmetry breaking and to the continuum theory.

During the last decade new selfdual solutions have entered the 
discussion. The aim is now to explain confinement in such a model 
via a detour through finite temperature, at $0 < T < T_{\rm dec}$. 
The new solutions are the Kraan-van Baal-Lee-Lu
\cite{Kraan:1998kp,Kraan:1998pm,Kraan:1998sn,Lee:1998bb} calorons with a 
general asymptotic holonomy 
${\cal P}_{\infty} \notin Z(N)$, not necessarily in the center of $SU(N)$.
For $SU(2)$ the asymptotic holonomy can be roughly identified with the 
real-valued spatial average of the Polyakov loop 
$L= (1/V) \sum_{\vec x} {\rm Tr} {\cal P}({\vec x})$.

Very recently, Diakonov and Petrov have worked out a model~\cite{Diakonov:2007nv} 
based on a gas of interacting caloron constituents, i.e. selfdual dyons, 
that offers already a complete picture of confinement at finite temperature. 
Although the presence of {\it both} selfdual dyons and antiselfdual antidyons 
has been ignored so far, the model is a convincing step forward. 
For the success of this description (which describes also the limit of low 
temperature !) the maximally nontrivial holonomy of the gauge field is crucial. 

In one paper~\cite{Gerhold:2006sk}, authored last year at Humboldt University, 
the capability of a caloron gas model to explain confinement has been explored 
in a Monte Carlo study. In this caloron model the opposite extreme case 
of dyons bound in calorons is dealt with. The importance of maximally nontrivial 
holonomy for the correct choice of the caloron solutions to be used in the model
was the central idea. Even a modest dissociation of calorons into 
slightly separated dyon-dyon pairs turned out sufficient to create a confining 
heavy-quark potential of the right order of magnitude. 

The assumptions of these models have to be confronted with the lattice.
Will we ever have the chance to confirm or disprove models of this type  
by analyzing generic Monte Carlo lattice configurations ? For some time already
our aim is to understand, albeit numerically, to what extent calorons and dyons 
coexist and are discernible in the Euclidean gauge fields, most probably below 
$T_{\rm dec}$. This has been the central question in our two previous lattice 
papers on the problem~\cite{Ilgenfritz:2004zz,Ilgenfritz:2006ju} and is the 
central question also now.

Traditionally, the presence of locally classical excitations like instantons
in Monte Carlo lattice gauge fields has been explored by using methods like 
cooling~\cite{Teper:1985rb,Ilgenfritz:1985dz,Hoek:1986nd,Polikarpov:1987yr}, 
restricted cooling~\cite{GarciaPerez:1998ru}, smearing~\cite{DeGrand:1997ss},
that replaced the more demanding RG cycling~\cite{DeGrand:1997gu,DeGrand:1997de}, 
and more general, by combinations of blocking and inverse 
blocking~\cite{DeGrand:1996ih,DeGrand:1996zb,Feurstein:1996cf}. 
In the result, a well-localized topological charge density (according to its 
field-theoretical definition~\cite{DiVecchia:1981qi,DiVecchia:1981hh}) becomes 
visible. All these methods actively change the gluonic field of the lattice 
configurations. Therefore they have been considered with some scepticism because 
of the methodical bias towards classical fields. Concerning our particular point 
of view, one might criticize that these methods also tend to hide the presence 
of dyonic constituents inside calorons.

For the evaluation of the topological charge density the situation has completely 
changed with the availability of methods dealing with overlap 
fermions~\cite{Neuberger:1997fp,Neuberger:1998wv} -- or other fermions with 
improved chiral properties~\cite{Gattringer:2000js,Gattringer:2000qu} -- as a 
probe to explore gauge fields. 
A definition~\cite{Niedermayer:1998bi,Hasenfratz:1998ri} of the topological charge 
density has been given involving the trace of the overlap Dirac operator. Then
ultraviolet filtering can be given a well-defined 
sense~\cite{Horvath:2002yn,Horvath:2002dw} by restricting the trace to the lowest 
fermionic eigenmodes with $|\lambda| < \lambda_{\rm cut}$ in the spectrum. 
The properties of this whole family of topological densities are strongly 
changing with $\lambda_{\rm cut}$ and have been investigated in detail in 
Ref.~\cite{Ilgenfritz:2007xu}.

In our two lattice papers~\cite{Ilgenfritz:2004zz,Ilgenfritz:2006ju} on the 
caloron/dyon issue we were relying on the smearing technique and using the 
field-theoretical definition of the topological charge density in terms of an 
improved lattice field-strength tensor~\cite{Bilson-Thompson:2002jk}. Additionally, 
in order to support the interpretation of the clusters of topological charge as 
calorons or dyons, the monopole content of the clusters has been analyzed in the 
maximal Abelian gauge.~\footnote{MAG was implemented on the lattice first in 
Refs.~\cite{Kronfeld:1987vd,Brandstater:1991sn}.}

With the present paper we return to the investigation of the caloron/dyon structure. 
We are replacing the field-theoretical topological charge density, which always 
requires smearing, by the overlap-based topological charge density in the 
ultraviolet filtered form mentioned above. We should remind the reader that chirally
improved lattice fermions~\cite{Gattringer:2000js,Gattringer:2000qu} (another 
realization of Ginsparg-Wilson~\cite{Ginsparg:1981bj} fermions) have already been 
used to analyze {\it unsmeared} lattice configurations for the presence of 
calorons~\cite{Gattringer:2002tg}. Before that, unimproved Wilson fermions have 
been employed~\cite{Ilgenfritz:2002qs} for the description of nearly classical 
calorons and dyons obtained by cooling. What was common to both techniques was 
inspired by the theoretically known behavior of zero modes of caloron-like 
configurations~\cite{Chernodub:1999wg}. Thus, particular emphasis was first given 
to the zero modes (or the real modes in case of the Wilson-Dirac operator), which 
must be present in configurations with topological charge $Q \ne 0$, and to the 
effect on them of changing the boundary conditions for the Dirac 
operator~\cite{Ilgenfritz:2002qs,Gattringer:2002tg,Gattringer:2003uq,Gattringer:2004pb}.
Confronting the zero-mode pattern with the picture revealed by smearing, it became 
clear~\cite{Gattringer:2003uq} that the zero modes are part of the 
topological structure of a typical Monte Carlo configuration but  
cannot exhaustively explain it.

In a recent paper~\cite{Bruckmann:2006wf} reporting a collaborative project of the 
Humboldt University and Regensburg University lattice groups, it has been described
how smearing and spectral filtering methods (with fermions and scalars) can be tuned
to each other as far as the topological charge density is concerned. In our present 
context the relation between the {\it fermionic} filtering and the result of 
smearing is relevant. In the parameter space of competing methods (number of modes 
vs. smearing steps) a mapping was defined by optimizing the cross-correlation 
between the respective topological charge densities. As just two extreme examples 
we quote the observations that 10 smearing steps are equivalent to the filtering by 
50 modes, while 20 smearing steps are equivalent to not more than 8 modes. These 
are only two arbitrarily taken cases of relatively mild and strong filtering. Of 
course, the structure changes (the number of lumps decreases) with increasing 
smearing steps. Moreover, even the parameter mapping does not guarantee that the 
clusters of the respective densities exactly coincide. The pointwise overlap amounts
only to 50 to 60 \%. Although in Ref.~\cite{Bruckmann:2006wf} chirally improved 
fermions~\cite{Gattringer:2000js,Gattringer:2000qu} were employed instead of overlap
fermions, the results give additional motivation and orientation for the present 
investigation and may be helpful to appreciate the new findings. We will explore 
the possibilities of identifying caloron-like and dyon-like structures for 
intermediate filtering employing 20 overlap eigenmodes.

What is the conjectured physical picture ? 
Our previous experience~\cite{Ilgenfritz:2004zz,Ilgenfritz:2006ju,Gerhold:2006sk} 
suggests, in accordance with the model of Diakonov and Petrov~\cite{Diakonov:2007nv}
that a ``plasma'' including calorons (with nontrivial holonomy) and dissolved dyonic
constituents may describe the field structure at $T < T_{\rm dec}$ rather well. 
It fails, however, to describe the essential features of lattice fields above 
$T_{\rm dec}$.
It has been guessed that calorons with intermediate holonomy~\footnote{The usual 
Harrington-Shepard calorons~\cite{Harrington:1978ve}, forming the basis of the
first nonperturbative description of finite-$T$ QCD~\cite{Gross:1980br}, represent 
the limiting case of trivial holonomy.} would describe the topological structure 
in the high-$T$ phase closely above $T_{\rm dec}$. A semiclassical evaluation of 
the path integral~\cite{Diakonov:2004jn} has shown, however, that the caloron 
becomes unstable against dissociation into dyons outside a narrow stability region 
$|L| > 0.72$. On the lattice, by suitable measures of 
selfduality~\cite{Gattringer:2002gn,Ilgenfritz:2007xu}, it has been 
observed that locally selfdual domains become suppressed above 
$T_{\rm dec}$. The topological susceptibility is known to slowly decrease (in the
case of $SU(2)$ gauge theory), and 
purely magnetic monopole excitations probably acquire an overwhelming importance.

In the confined phase the caloron model indeed describes confinement, even if the 
dissociation of calorons into dyons remains incomplete and within a description 
by a phenomenological choice of the $\rho$ distribution. The size variable 
$\rho^2=d/(\pi T)$ in the caloron case represents a natural extension of the 
size parameter $\rho^2$, usually assigned to the (Euclidean) spherical lumps of 
action seen at $T\to 0$, to higher temperatures in the confinement phase when 
the distance $d$ between the constituents may be 
$d \sim 1/(\pi T)$ or bigger. Dissociation is 
increasing with rising temperature towards $T_{\rm dec}$.~\footnote{This  
tendency is also supported by the cooling results in Ref.~\cite{Ilgenfritz:2004ws}.} 
This is in agreement with a practically temperature independent 
$\rho$-distribution for $T < T_{\rm dec}$, similar to the usual instanton 
parametrization~\cite{Gerhold:2006sk}. Let us remark that at very low temperature 
the carriers of topological charge become more and more difficult to distinguish 
from genuine instantons  by means of gauge-invariant observables alone (action and 
topological charge density). All this has lead to the conjecture that close to the 
deconfining phase transition the dyonic content of calorons becomes maximally 
manifest. Therefore we concentrate here first on this temperature. 

In the present investigation we have dispensed with ({\it i}) the use of smearing 
for the detection of clusters of topological charge, and with ({\it ii}) the 
maximally Abelian gauge needed to determine the monopole content of the latter. 
We gave up, on the other hand, the exclusive focus on the zero 
mode(s)~\cite{Gattringer:2002tg,Gattringer:2004pb} of the configurations being 
under investigation. We have concentrated instead on the effect of changing the 
fermionic boundary conditions on the whole overlap-based topological charge density 
mapped out by a given (not too large) number of eigenmodes. 
The dependence on boundary conditions has not yet been systematically investigated. 
Our present paper is a first step in this direction and hopefully stimulates such 
a thorough investigation. Also concerning the $SU(2)$ gauge theory, this paper is 
the first application of the overlap-based topological charge density. For two 
colors it suffices to restrict oneself to the simultaneous consideration of periodic
vs.  antiperiodic boundary conditions. We shall see that the mode-truncated 
topological charge densities corresponding to the two boundary conditions, completed
by the local Polyakov loop, allow us to classify the visible topological charge 
clusters as calorons and separate dyons, respectively.

The paper is organized as follows.
In section~\ref{sec:spectrum_action} we define the technical details of our
analysis: the action~\cite{Luscher:1984xn,Curci:1983an}, the overlap Dirac 
operator~\cite{Neuberger:1997fp,Neuberger:1998wv} and the corresponding topological 
charge density~\cite{Niedermayer:1998bi,Hasenfratz:1998ri}. Under conditions of 
confinement, but close to the transition temperature, we critically check the 
stability of the fermionic topological charge (given by the index of the overlap 
Dirac operator) and the continuity of the low-lying spectrum under smearing and 
with respect to a change of fermionic boundary conditions. This check forces us 
to abandon the standard Wilson gauge action and motivates the choice of the 
L\"uscher-Weisz action that successfully passes the test. The bulk of investigations
is performed using the tadpole-improved L\"uscher-Weisz 
action~\cite{Gattringer:2001jf}. In section~\ref{sec:transition_point} we determine 
the critical $\beta_{{\rm imp},c}$ for the quenched thermal phase transition 
with this action. The search for the transition is restricted to a $20^3 \times 6$ 
lattice that will be used in the following. Next, in section~\ref{sec:overlap}, we 
explain how the mode-truncated, overlap-based topological charge density obtained 
with the two different fermionic boundary conditions can be used to extract calorons
and dyons from unsmeared configurations, but restricted to the resolution given 
by the number of eigenmodes. This is practised for an ensemble generated on 
top of the phase transition. In the future we hope to proceed with this analysis
deeper into the confinement and the deconfinement phases. A discussion of the results
in the light of related work, our conclusions and an outlook will be presented 
in section~\ref{sec:conclusion}.

\section{Wilson vs. L\"uscher-Weisz action: the stability of the
Dirac spectrum}
\label{sec:spectrum_action}

\subsection{The action}
\label{subsec:action}

We employ for the actual analysis of the caloron/dyon content of $SU(2)$ gauge
theory the tadpole improved action of the L\"uscher-Weisz 
form~\cite{Alford:1995hw,Bornyakov:2005iy}
\begin{equation}
S =   \beta_{\rm imp} \sum_{\rm pl} S_{\rm pl}
        - {\beta_{\rm imp} \over 20 u_0^2} \sum_{\rm rt} S_{\rm rt} \; ,
\label{eq:improved_action}
\end{equation}
where $S_{\rm pl}$ and $S_{\rm rt}$ denote plaquette and $1 \times 2$ rectangular
loop terms in the action,
\begin{equation}
S_{\rm pl,rt}\ = \ \frac{1}{2}~{\rm Tr}~(1-U_{\rm pl,rt}) \, .
\label{eq:terms}
\end{equation}
The parameter $u_0$ is the input tadpole improvement factor taken here equal to 
the fourth root of the average plaquette
$W_{1\times1}=\langle (1/2)~{\rm Tr}~U_{\rm pl} \rangle$. 
For $SU(2)$ gauge theory, the tadpole factor $u_0$ has been selfconsistently 
determined first in Ref.~\cite{Bornyakov:2005iy} for a few $\beta_{\rm imp}$ 
values in the case of vanishing temperature on $L^4$ lattices with a suitable 
lattice size $L$ for each value of the bare coupling constant. The result is 
given in Table I. For the convenience of the reader and later reference to 
the lattice scales we present the corresponding values of the string tension 
in lattice units.

\begin{table}[htb]
\begin{center}
\caption{Details of the simulations with tadpole-improved L\"uscher-Weisz
action at $T=0$}
\vspace{.3cm}
\setlength{\tabcolsep}{0.55pc}
\begin{tabular}{cccccc}
$\beta_{\rm imp}$ & L & $u_0$ & $<P>^{1/4} $ & $\sqrt{\sigma a^2}$ \\
\hline
2.7 & 12 &  0.87164 & 0.87165(2)  & 0.60(5)   \\
3.0 & 12 &  0.89485 & 0.89478(2)  & 0.366(8) \\
3.1 & 12 &  0.90069 & 0.90069(1)  & 0.309(6)  \\
3.2 & 16 &  0.90578 & 0.905765(3) & 0.258(5)  \\
3.3 & 16 &  0.91015 & 0.910152(4) & 0.219(3)  \\
3.4 & 20 &  0.91402 & 0.914020(2) & 0.180(3)  \\
3.5 & 20 &  0.91747 & 0.917481(1) & 0.151(3)  \\
\hline
\end{tabular}
\label{tab:tab1}
\end{center}
\end{table}

In our simulations we have not included one-loop corrections to the coefficients 
nor considered non-planar 6-link loops the coefficient of which would be purely 
perturbative. We have adopted the $u_0$ values obtained at zero temperature also 
for the simulations at $T \approx T_{\rm dec}$.

\subsection{The overlap Dirac operator}
\label{subsec:overlap_operator}

The overlap Dirac operator is a particular solution of the Ginsparg-Wilson 
relation~\cite{Ginsparg:1981bj}
\begin{equation}
D \gamma_5 + \gamma_5 D = \frac{a}{\rho} D \gamma_5 D \; , 
\label{eq:GW_relation}
\end{equation}
where $\rho=O(1)$ is a dimensionless parameter not to be 
confused~\footnote{We prefer to keep this standard notation.} 
with the instanton or caloron size parameter above. These operators 
have nearly perfect chiral properties. In particular, the Atiyah-Singer index 
theorem is fulfilled at finite lattice spacing, with $N_{\pm}$ clearly recognizable 
zero modes with positive or negative chirality related to the topological charge
\begin{equation}
Q_{\rm index} = N_{-} - N_{+} \; .
\end{equation}
This is unambiguous as long as the configurations satisfy certain weak smoothness 
requirements~\cite{Hernandez:1998et}. A Neuberger operator can be constructed 
starting from an arbitrary input Dirac operator (with bad chiral symmetry or 
with already improved chiral symmetry) through the steps we describe now. In our 
case, we take as the input kernel the simple Wilson-Dirac operator. In this case, 
the emerging Neuberger overlap operator~\cite{Neuberger:1997fp,Neuberger:1998wv} is
\begin{equation}
D_{\rm ov} = \frac{\rho}{a}~\left( 1 + D_{\rm W}/\sqrt{D_{\rm W}^{\dagger}~D_{\rm W}} \right)
\; , \qquad D_{\rm W} = M - \frac{\rho}{a} \; ,
\label{eq:Neuberger_operator}
\end{equation}
where $D_{\rm W}$ is the Wilson-Dirac operator with a negative mass term $\rho/a$.
$M$ is the Wilson hopping term with $r=1$. An optimal choice is $\rho \approx 1.4$ .
By construction, the operator $D_{\rm ov}$ satisfies the Ginsparg-Wilson relation.
In order to compute the sign function in the alternative expression 
\begin{equation}
D_{\rm W}/\sqrt{D_{\rm W}^{\dagger}~D_{\rm W}} = \gamma_5~\mathrm{sgn}\left(H_{\rm W}\right) , \qquad
H_{\rm W} = \gamma_5~D_{\rm W} \; , 
\end{equation}
we have used the minmax polynomial approximation~\cite{Giusti:2002sm}. Furthermore, 
the low-mode projection has been used: 80 eigenmodes of the hermitean Wilson-Dirac 
operator $H_{\rm W}$ have been treated explicitely.

The topological charge density can be expressed in the form~\cite{Niedermayer:1998bi}
\begin{equation}
q(x) = - {\rm tr} \left[ \gamma_5 \left( 1 - \frac{a}{2\rho} D_{\rm ov}(x,x) \right)\right] \; ,
\label{eq:full_density}
\end{equation}
where ${\rm tr}$ denotes the trace only over color and spinor indices. This form of 
the topological charge density contains vacuum fluctuations of all scales. In the
apparent chaos remarkable global, low-dimensional 
structures~\cite{Horvath:2002zy,Horvath:2003yj,Horvath:2004gw} are formed. 
They are three-dimensional at the percolation threshold~\cite{Ilgenfritz:2007xu}.
The ultraviolet filtered (mode-truncated) 
density~\cite{Horvath:2002yn} is written as a truncated sum over $\lambda$ as
the dimensionless eigenvalues of $a~D_{\rm ov}/\rho$,
\begin{equation}
q_{\lambda_{\rm cut}}(x) = - \sum_{|\lambda| \le \lambda_{\rm cut}} \left( 1 - \frac{\lambda}{2} \right) \psi^{\dagger}_{\lambda}\gamma_5 \psi_{\lambda}(x)  \; ,
\label{eq:truncated_density}
\end{equation}
however, shows clustering of topological charge~\cite{Ilgenfritz:2007xu} in 
four-dimensionally coherent clusters similar~\cite{Bruckmann:2006wf} to structures 
usually revealed by smoothing the gauge field. This is the level of resolution
where calorons and dyons may appear (or not).

The integral over $q_{\lambda_{\rm cut}}(x)$ gives $Q$ corresponding to the 
Atiyah-Singer index theorem, independent of $\lambda_{\rm cut}$, because only the 
zero modes contribute to $Q$ according to their chirality. It is remarkable, but 
generally observed for the overlap Dirac operator, that if there are zero modes 
within a configuration, they all have the same chirality.

\subsection{The Dirac spectrum under smearing and varying boundary conditions}
\label{subsec:smearing}

The cross-relation between topological charge density and local structure of 
the Polyakov loop is typical for calorons and their constituents. In order 
to map out the Polyakov loop we need a modest amount of APE smearing in this study. 
A second role of smearing in our present context is that we want to monitor 
the independence of the index of the overlap Dirac operator and a smooth 
dependence of the low lying spectrum on the number of APE smearing steps.
We regard this as a necessary prerequisite that this part of the spectrum reflects
medium-scale and infrared properties only. Thus, a minimal requirement for the 
Dirac operator as well as for the lattice action (to prevent lattice artefacts 
that could give rise to unphysical zero modes) is the continuity of the spectrum 
under moderate smearing. This in fact selects admissible actions and an admissible 
range of the respective coupling.

Smearing is an iterative sequence of four-dimensional link substitutions, 
where links are replaced by a weighted average of the links and the staples 
$$U_\mu^\nu(x)=U_\nu(x)U_\mu(x+\hat{\nu})U_\nu^\dagger(x+\hat{\mu})$$ 
surrounding it:
\begin{equation}
U_\mu(x)\to\mathcal{P}\big[(1-\alpha) U_\mu(x)+\frac{\alpha}{6}\sum_{\nu \neq \mu} 
\left( U_\mu^\nu(x) + U_\mu^{-\nu}(x) \right) \big]\,.
\end{equation}
Here $\mathcal{P}$ denotes the projection onto the gauge group. For $SU(2)$ this 
is just a rescaling of the matrix by a scalar. We choose the smearing parameter
$\alpha=0.45$ following \cite{DeGrand:1997ss} where an optimal smearing schedule 
has been searched for. Within some limits, this parameter could be traded against 
the number of smearing steps. We allow for $N_{\rm APE} \leq 10$ iterations. 

\begin{figure*}[!htb]
\begin{center}
\includegraphics[width=.7\textwidth]{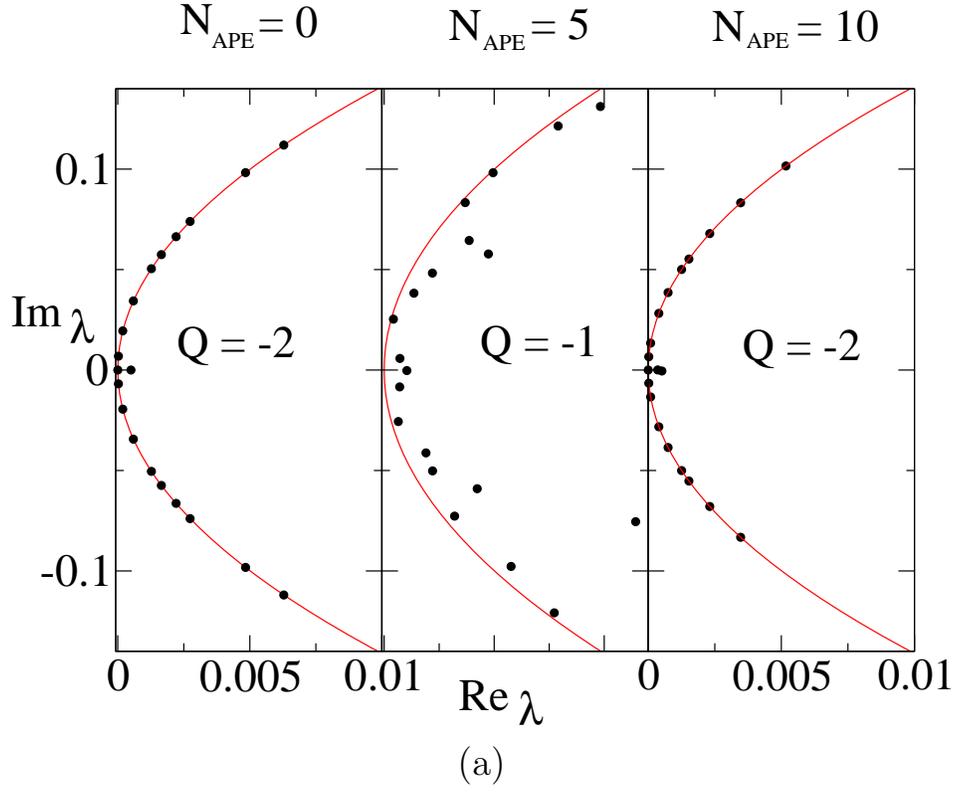}\\
\Large{(a)}\\
\vspace{1cm}
\includegraphics[width=.7\textwidth]{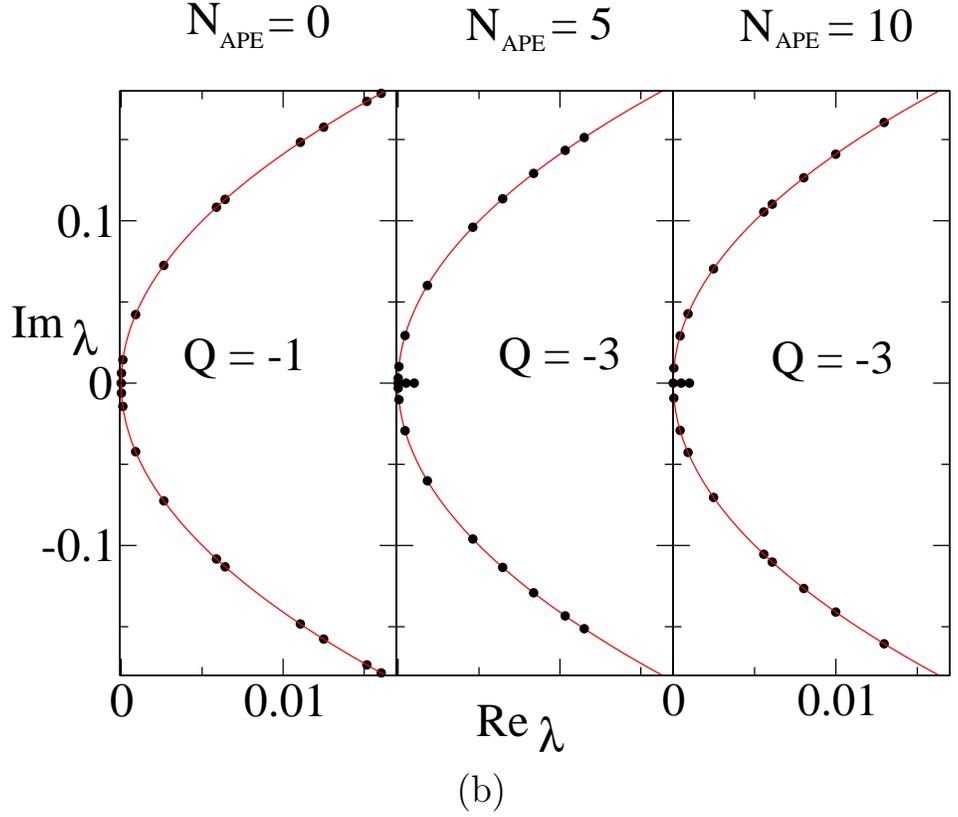}\\
\Large{(b)}\\
\end{center}
\caption {The eigenvalue spectra of the Dirac overlap operator 
with (a) periodic and (b) antiperiodic temporal boundary conditions
for equilibrium ($N_{\rm APE}=0$) and smeared ($N_{\rm APE}=5$ and 10) 
pure $SU(2)$ gauge configurations on a $20^3\times 6$ lattice, 
generated with the standard Wilson action at $\beta = 2.40$ .}
\label{fig:Wilson_spectra}
\end{figure*}

It is known since long~\cite{Gockeler:1989qg} that the Wilson action is 
problematic with respect to dislocations. We remind the reader that the 
definition of a dislocation depends {\it both} on the action in use for 
the generation of gauge configurations {\it and} on the prescription chosen 
to define the topological charge (or charge density). 
In the case of the $SU(2)$ Wilson action together with the (geometrical) 
Phillips-Stone topological charge~\cite{Phillips:1986qd}, scaling of the 
topological susceptibility~\cite{Kremer:1988ww} was an unsuspected fact 
before Pugh and Teper~\cite{Pugh:1989qd} have shown that the observed 
value of $\langle Q^2 \rangle$ was dominated by excitations of size 
$O(a)$ that would not survive blocking. The lesson we draw from this example
is that the bosonic topological charge 
\begin{equation}
Q_{\rm gluon}= \int d^4x~q_{\rm gluon}(x) 
\end{equation}
to be assigned to a configuration by a suitably improved gluonic topological 
charge density
\begin{equation}
q_{\rm gluon}(x) = \frac{g^2}{32~\pi^2} \varepsilon_{\mu\nu\rho\sigma} 
{\rm Tr} \left(F_{\mu\nu} F_{\rho\sigma} \right) \; ,
\end{equation}
and a geometrically defined version of the topological 
charge~\cite{Phillips:1986qd,Luscher:1981zq} may be in systematical disagreement 
due to the presence of lattice artefacts. This can be the case even though it
might not appear as a scaling violation of the topological susceptibility. 
In a first application of the overlap operator to $SU(2)$ gauge fields generated 
with Wilson action~\cite{Gubarev:2005az}, a reasonable continuum limit of the 
susceptibility $\langle Q^2 \rangle/V$ with $Q$ defined via the index of the 
overlap operator has been found. Therefore the Wilson action was not suspicious.
The value of $\chi_{\rm top}$, however, was somewhat large compared 
to other estimates for the $SU(2)$ gauge theory.~\footnote{The scaling property 
of the topological susceptibility would only be lost, resulting in a diverging 
susceptibility in the continuum limit, if local excitations would exist, that 
give rise to highly localized zero modes and would have a Wilson action 
less than $S_W < \frac{12}{11}\pi^2$~\cite{Pugh:1989ek}.} 

The field-theoretic definition of the topological charge density that we have 
used in our previous papers~\cite{Ilgenfritz:2004zz,Ilgenfritz:2006ju} employs  
the improved field strength tensor~\cite{Bilson-Thompson:2002jk}.  
The geometrical definition of the topological charge $Q$ of a configuration is 
replaced in our present context by the index of the overlap Dirac operator,
and the topological density by the corresponding expression (\ref{eq:full_density})
given above.  At this point, checking the above-formulated requirements, disturbing 
features of the Wilson action are encountered. At first, the roughness of the 
configurations results in a relatively bad performance of the ARPACK package 
used to diagonalize the overlap Dirac operator. This probably leads to a bad 
reproducibility of the measured index. Thus, the latter can easily be misidentified 
due to zero modes pinned to dislocations. During the first smearing steps such 
dislocations become even more singular. 

As a typical example we show in 
Fig.~\ref{fig:Wilson_spectra} the lowest 20 eigenvalues according to the two 
boundary conditions imposed, without smearing and with 5 and 10 smearing steps, 
for a $20^3\times6$ configuration generated with the Wilson action at $\beta=2.4$ .
This $\beta$ was chosen below the critical value $\beta_c(N_{\tau}=6)=2.4265(30)$ 
reported for the Wilson action in Ref.~\cite{Fingberg:1992ju}. 
We see jumps of the measured topological charge, $|\Delta Q_{\rm index}|=1$, 
between subsequent stages of smearing and occurring under a change of the boundary 
conditions (temporally periodic vs. antiperiodic).
During the first steps, smearing changes only the short range
structure. The changing index counts here essentially the dislocations.
Hence, the number of zero modes rapidly changes with the APE smearing steps. 

\begin{figure*}[!htb]
\begin{center}
\includegraphics[width=.7\textwidth]{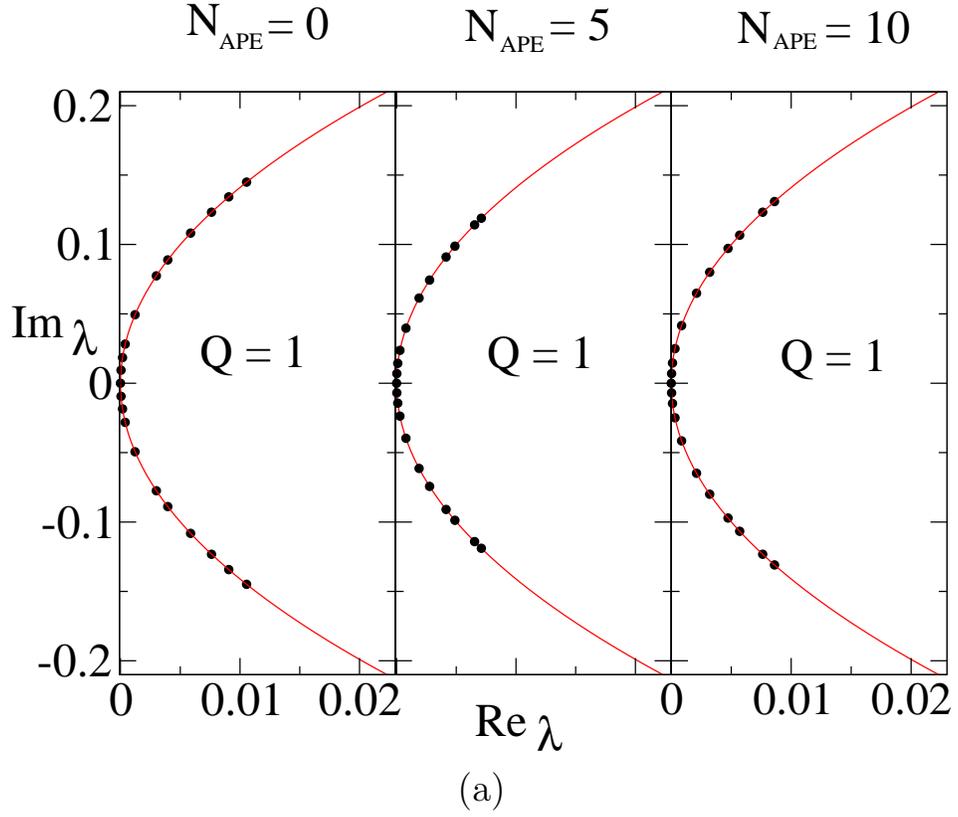}\\
\Large{(a)}\\
\vspace{1cm}
\includegraphics[width=.7\textwidth]{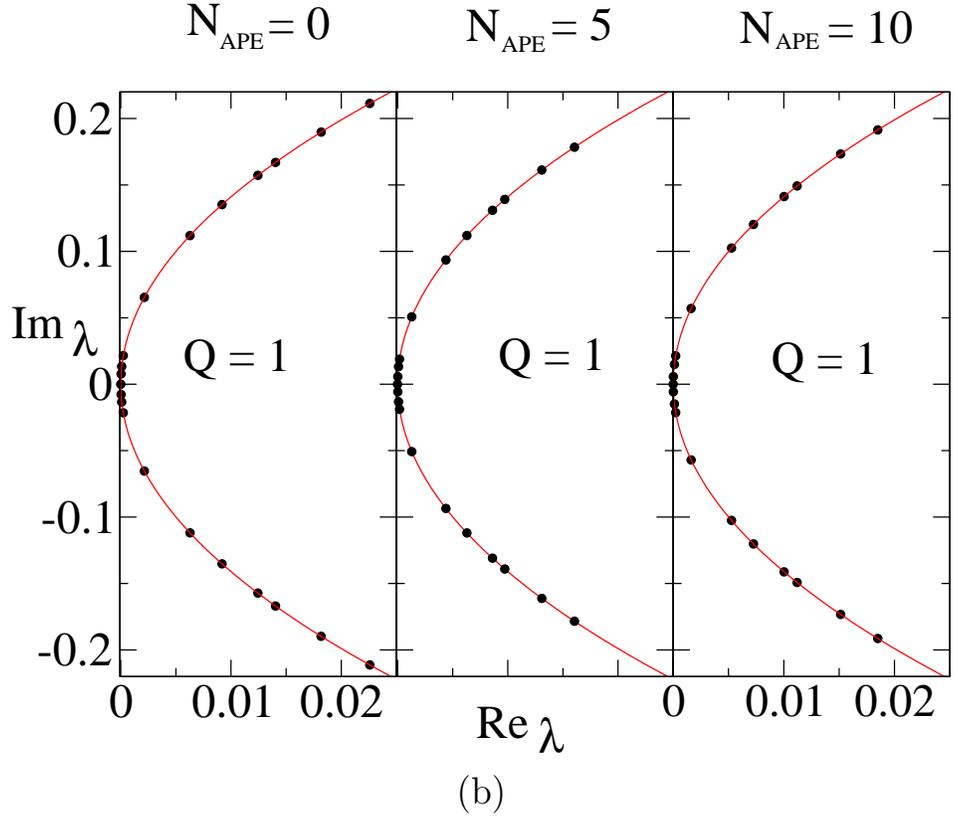}\\
\Large{(b)}\\
\end{center}
\caption {The eigenvalue spectra of the Dirac overlap operator 
with (a) periodic and (b) antiperiodic temporal boundary conditions
for equilibrium ($N_{\rm APE}=0$) and smeared ($N_{\rm APE}=5$ and 10) 
pure $SU(2)$ gauge configurations on a $20^3\times 6$ lattice, 
generated with the tadpole-improved L\"uscher-Weisz action at 
$\beta_{\rm imp} = 3.20$ .}
\label{fig:Luescher_Weisz_spectra}
\end{figure*}

The fact that the L\"uscher-Weisz action is advantageous to facilitate our study,
has been confirmed for a number of $\beta_{\rm imp}$ values. The result is  
demonstrated in Fig.~\ref{fig:Luescher_Weisz_spectra} for a typical configuration 
from a L\"uscher-Weisz ensemble at $\beta_{\rm imp}=3.2$. The number of zero modes 
is independent of the type of temporal boundary condition and does not change with
the number of APE smearing steps (as long as smearing is moderate, 
say $N_{\rm APE} \le 10$). 
For $\beta_{\rm imp} \geq 3.2$ we have never encountered such ambiguities as 
seen in the Wilson case. In Section~\ref{sec:transition_point} we will see that 
the critical inverse gauge coupling for this action is 
$\beta_{{\rm imp},c}(N_{\tau}=6)=3.248(2)$. The successful check presented in 
Fig.~\ref{fig:Luescher_Weisz_spectra} has been performed for a situation close 
but clearly below the phase transition.

We should stress, however, that configurations created by means of the 
L\"uscher-Weisz action may also turn out too ``rough'' at sufficiently
low $\beta_{\rm imp}$ values. For example, exploring the temperature range 
around $T_{\rm dec}$ on a coarser lattice with $N_{\tau}=4$ (i.e. at lower 
$\beta_{\rm imp}$), we found that the described ambiguities reappear.

A surprising observation in the case of both actions is that the interval 
covered by the 20 lowest eigenvalues does not systematically expand under 
the application of smearing steps. This differs from the behavior seen in 
Ref.~\cite{Gattringer:2006wq} for $16^4$ configurations generated with the 
(tree-level) L\"uscher-Weisz action. The spectrum there was considered not 
for the overlap Dirac operator but for the chirally improved Dirac operator 
proposed in Ref.~\cite{Gattringer:2000js,Gattringer:2000qu}. It would be 
interesting to compare the two Dirac operators in their behavior  
under smearing for different lattice ensembles (provided the smearing-induced 
changes are smooth).

\section{Locating the finite temperature phase transition}
\label{sec:transition_point}

The last observations make clear that we need to choose $N_{\tau} \geq 6$
for the purpose of this investigation. Let us now look for a more precise 
location of the deconfinement transition. On the $20^3\times6$ lattice, 
varying $\beta_{\rm imp}$, we have studied the behavior of the Polyakov loop 
and of the Polyakov loop susceptibility. We used a polynomial fit for $u_0$ 
as a function of $\beta_{\rm imp}$, based on the measured values shown in 
Table I, in order to provide the corresponding tadpole improvement factor for 
each simulation point $\beta_{\rm imp}$. We stress again that this non-perturbative 
determination, strictly speaking, is well-established only for temperature $T=0$.

\begin{figure*}[!htb]
\begin{center}
\includegraphics[width=.75\textwidth,angle=0]{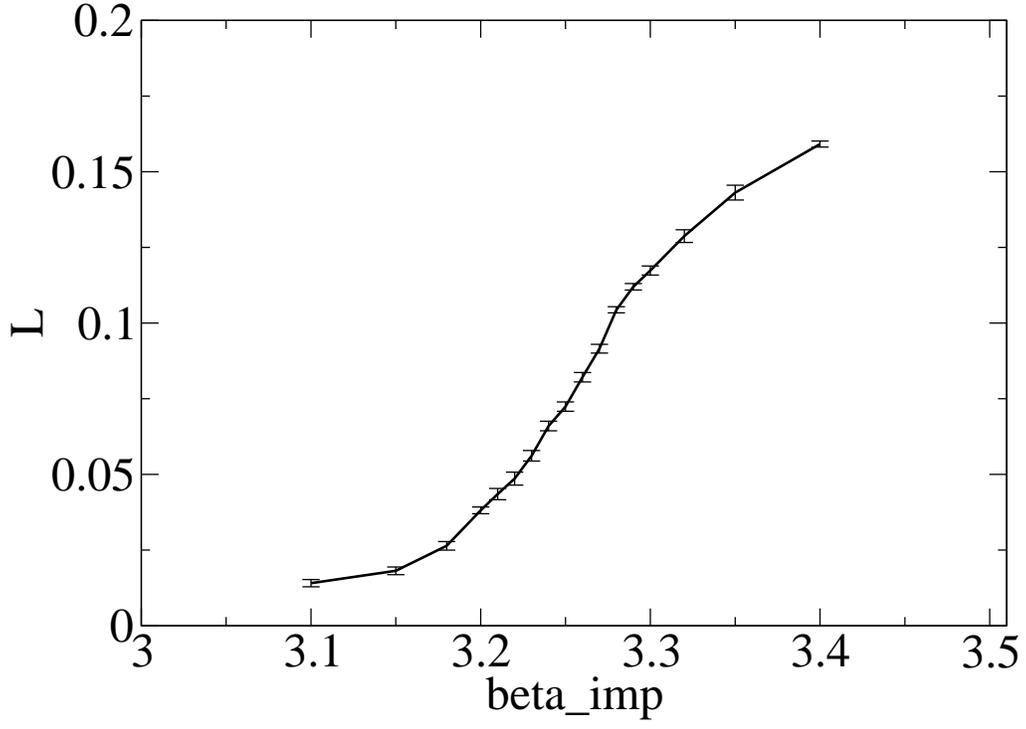}\\
\Large{(a)}\\
\vspace{2.0cm}
\includegraphics[width=.75\textwidth]{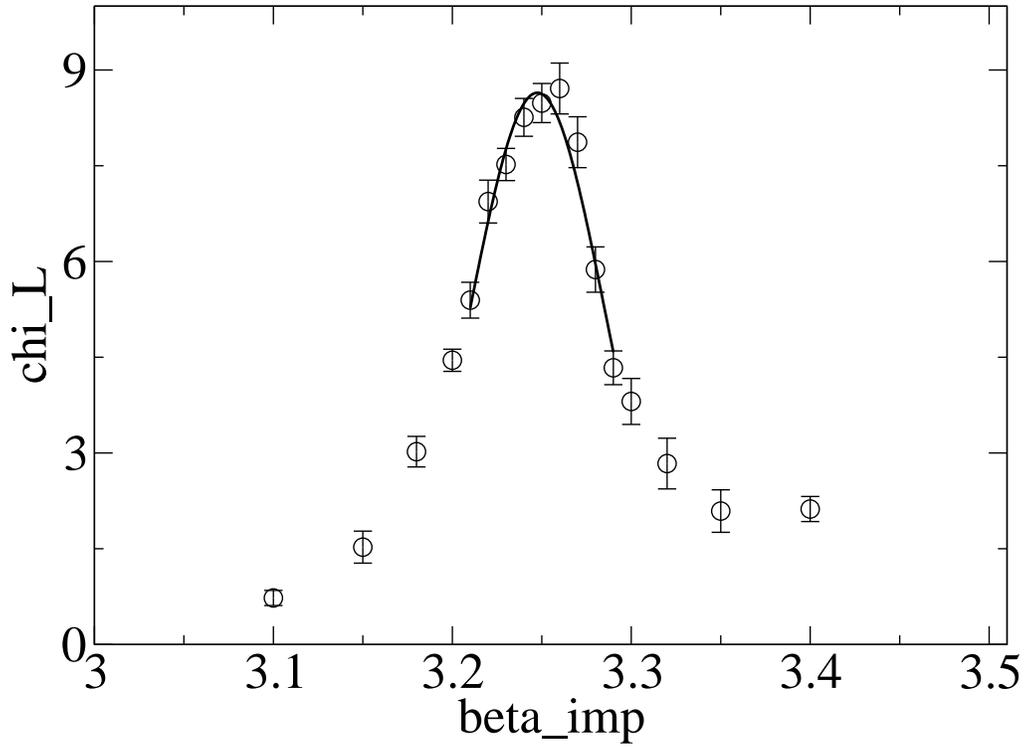}\\
\Large{(b)}\\
\end{center}
\caption {(a) The ensemble average of the modulus $|L|$ of the average
Polyakov loop
$L=\frac{1}{V}\sum_{\vec x} {\rm Tr} {\cal P}({\vec x})$ 
as function of $\beta$ for the 
tadpole-improved L\"uscher-Weisz action on a $20^3\times 6$ lattice. 
(b) The susceptibility of $|L|$ as function of $\beta$.}
\label{fig:Polyakovloop_average_suscept}
\end{figure*}

We have measured the Polyakov loop and its susceptibility in the range
from $\beta_{\rm imp}=3.1$ to $3.4$ with different statistics per data
point. The simulation data between $\beta_{\rm imp}=3.20$ and 3.29, 
in the immediate vicinity of the phase transition, have been collected 
in 100,000 to 300,000 Monte Carlo sweeps per $\beta_{\rm imp}$ value
while the Polyakov loop 
$L= (1/V) \sum_{\vec x} {\rm Tr} {\cal P}({\vec x})$
was measured after every sweep. In the closer vicinity of the phase 
transition we have fitted the susceptibility data by a Gaussian. The data 
and the fit of the susceptibility are presented in 
Fig.~\ref{fig:Polyakovloop_average_suscept}.
For the determination of the errors, the blocked jackknife method was 
used with a block size of 2000 measurements. From the fit we are able to 
locate the deconfinement transition at $\beta_{{\rm imp},c}=3.248(2)$ for 
$N_{\tau}=6$.  This confirms our preliminary choice made in 
Sect.~\ref{sec:spectrum_action} of $\beta_{\rm imp}=3.2$ for a check of
smoothness of the overlap Dirac operator that should be done in the confinement 
phase on a lattice of the same size. Interpolating the data in Table I
we estimate $\sqrt{\sigma}a=0.236(5)$ at $\beta_{{\rm imp~}c}$ corresponding
to $T_{\rm dec}/\sqrt{\sigma}=0.71(2)$. 

\section{Finding calorons and dyons using periodic and antiperiodic modes}
\label{sec:overlap}

In the Introduction we have argued why we should first search 
on top of the phase transition for calorons with nontrivial holonomy 
and why we anticipate to find them partly separated into dyons.
We have chosen $\beta_{\rm imp}=3.25$ very close to the transition point
for the following study of topological charge clustering. 
Our analysis is based on $20$ lowest-lying modes for an ensemble of $O(20)$ 
quenched configurations at the deconfinement transition. This was a realistic
task within the capability of a standard modern PC within a few weeks. 

The topological charge density of an equilibrium Monte Carlo field configuration 
is represented by the mode-truncated, i.e. ultraviolet filtered, topological 
charge density (\ref{eq:truncated_density}). In that definition the temporal 
boundary condition was not specified, that should be applied in the construction 
of the Wilson-Dirac and the Neuberger overlap operator 
(\ref{eq:Neuberger_operator}). From the work of Gattringer and 
Schaefer~\cite{Gattringer:2002tg} we know that the single zero mode of a 
$Q=\pm1$ Monte Carlo configuration eventually hops between $N_{\rm color}$ 
positions. On the other hand, the topological charge density of a (quenched) 
lattice configuration cannot depend on the purely analysing fermions, in particular
not on the boundary conditions imposed to them. The most suggestive rule for the 
topological charge density, if given by the zero-mode part of 
(\ref{eq:truncated_density}) alone, would be to average over the boundary 
conditions, which eventually (but not always !) lead to a different localization 
of the zero mode. This recipe is now applied to the topological charge density 
with the inclusion of the low-lying non-zero modes, too.

\begin{figure*}[!htb]
\begin{center}
\includegraphics[width=.4\textwidth]{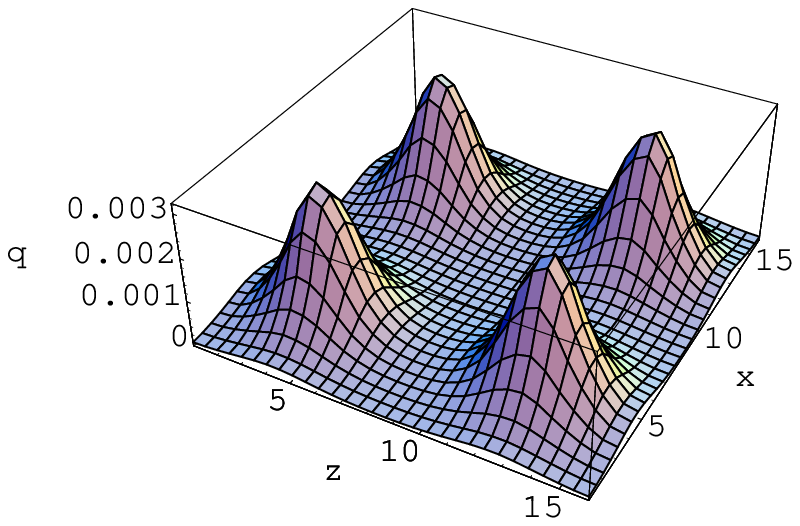}%
\hspace{1cm}\includegraphics[width=.4\textwidth]{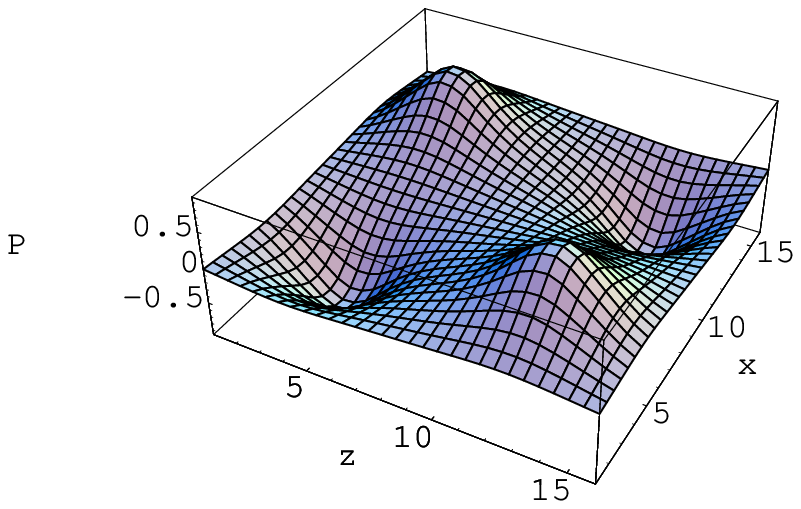}\\ 
\Large{(a)}\\
\vspace{1cm}
\includegraphics[width=.4\textwidth]{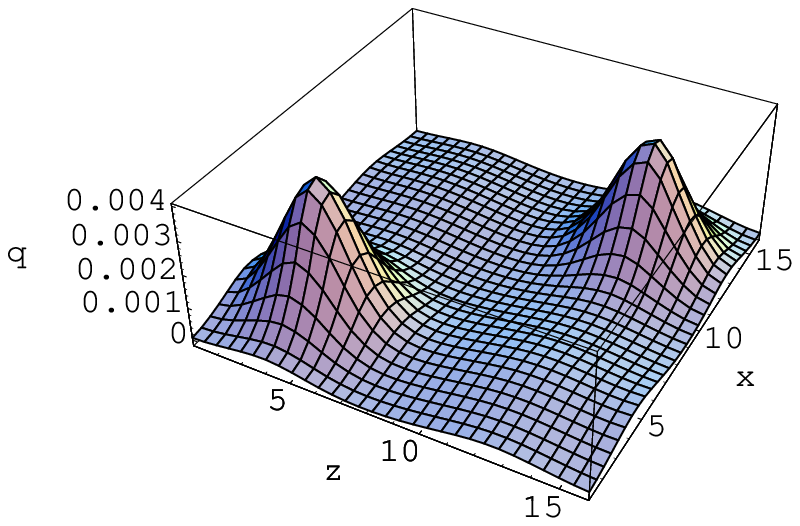}%
\hspace{1cm}\includegraphics[width=.4\textwidth]{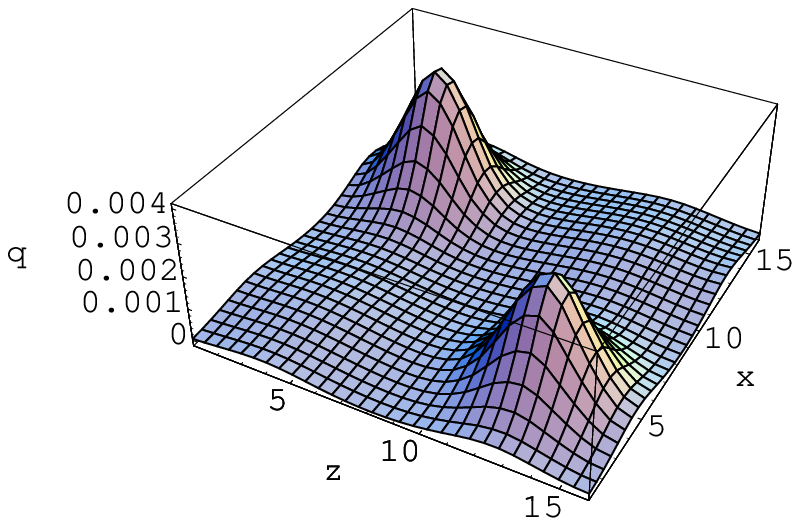}\\
\Large{(b)}\\
\end{center}
\caption {(a) The gluonic topological charge density $q_{\rm gluon}(x)$ (left) 
and the Polyakov 
loop $p({\vec x})$ (right) for a classical $Q=2$ configuration generated at 
maximally nontrivial holonomy (asymptotically $p({\vec x})=0$) on a $16^3*4$ 
lattice (with four dyons maximally separated in 
the $(x,z)$-plane). (b) The fermionic topological charge density $q^{(p/a)}(x)$ 
reconstructed 
out of the 20 lowest eigenmodes of the overlap Dirac operator with periodic 
(right) and antiperiodic (left) temporal boundary conditions.}
\label{fig:Q=2-caloron}
\end{figure*}

We illustrate this in Fig.~\ref{fig:Q=2-caloron} for a classical charge $Q=2$ 
caloron solution with nontrivial holonomy in a state of maximal separation 
into four dyons. The upper panels show the gluonic definition $q_{\rm gluon}(x)$ 
of the topological charge density and the profile of the Polyakov loop 
$p({\vec x}) = (1/2)~{\rm Tr} {\cal P}({\vec x})$ over a two-dimensional section of 
a $16^3 \times 4$ lattice. 
The gluonic topological charge density recognizes all the four constituents as 
positive peaks while the Polyakov loop distinguishes the constituents according 
to the local holonomy, i.e. positive and negative values of the Polyakov loop.
In the fermionic definition of the topological charge density, 
$q_{\lambda_{\rm cut}}(x)$, we content ourselves to only 20 lowest modes. 
We find that this filtered density depends on the boundary condition $b$, with 
$b=p$ denoting periodic and $b=a$ denoting antiperiodic temporal boundary conditions.
The charge densities present a different profile depending on the type of boundary 
conditions. The antiperiodic boundary condition highlights the constituents with 
negative local Polyakov loop, whereas the periodic boundary condition emphasizes 
the complementary constituents with positive local Polyakov loop. 
The ``true'' topological charge density (that is well-represented by the gluonic 
definition in this classical case) is well approximated by an average of the two 
fermionic topological charge density functions 
$q^{(p)}_{\lambda_{\rm cut}}(x)$ and $q^{(a)}_{\lambda_{\rm cut}}(x)$,
\begin{equation}
q^{(b)}_{\lambda_{\rm cut}}(x) = - \sum_{|\lambda_{b}| \le \lambda_{\rm cut}} \left( 1 - \frac{\lambda_{b}}{2} \right) \psi^{(b)\dagger}_{\lambda_{b}} \gamma_5 \psi^{(b)}_{\lambda_{b}}(x)  \; ,
\label{eq:truncated_density_bc}
\end{equation}
where the superscript $b=p$ or $b=a$ of the modes (the subscript of the eigenvalues)
refers to the boundary condition.
\begin{figure*}[!htb]
\begin{center}
\includegraphics[width=.7\textwidth]{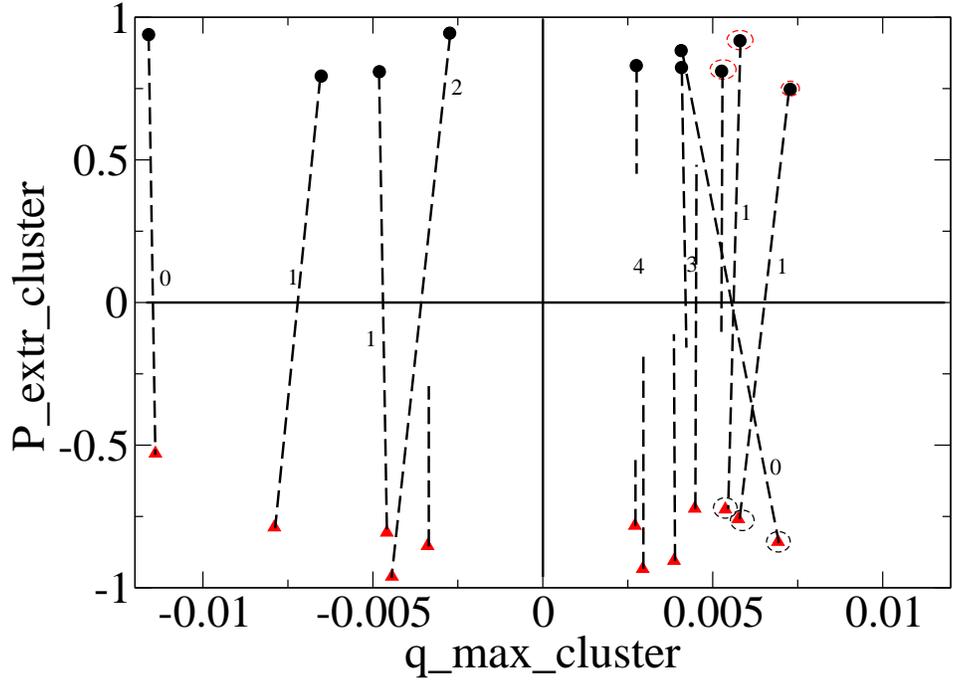} \\
\Large{(a)}\\
\vspace{1.5cm}
\includegraphics[width=.7\textwidth]{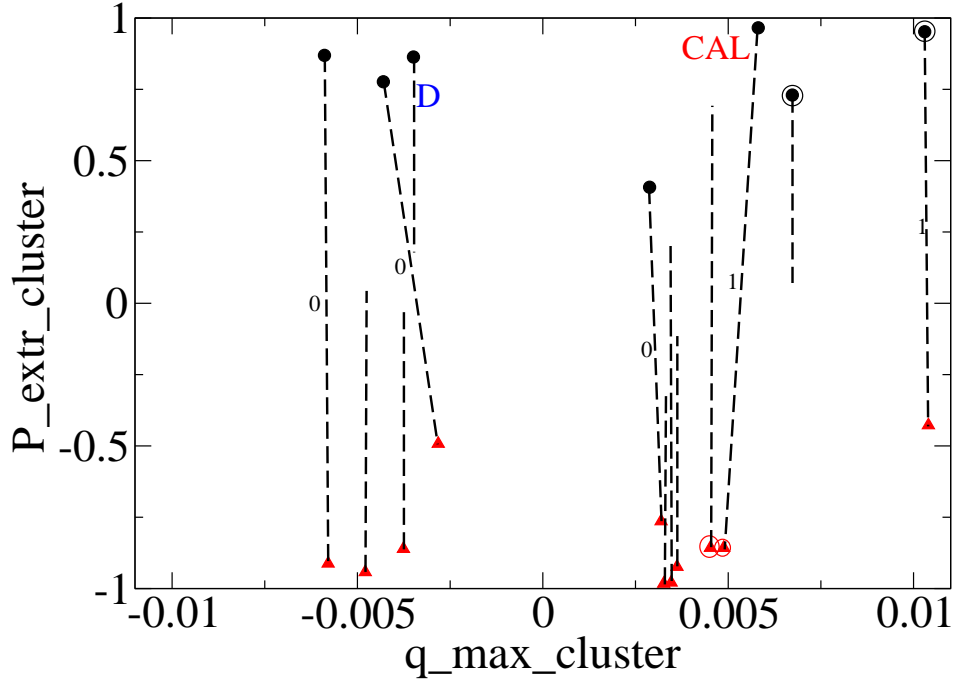} \\
\Large{(b)}\\
\end{center}
\caption{The maxima of clusters of the fermionic $|q(x)|$ seen under 
periodic boundary condition (filled circles) and under antiperiodic boundary 
condition (filled triangles) for two configurations (a) and (b) in the sample, 
shown in the $(q_{\rm max~cluster},P_{\rm extr~cluster})$ plane 
(the precise meaning is explained in the text).
Peaks at opposite-sign of $P_{\rm extr~cluster}$, that are connected 
by dashed lines, have appeared under different boundary conditions at 
the same space-time position (``not jumping'') and are interpreted as 
calorons. Isolated peaks have appeared only once under the respective 
boundary condition at the given position (``jumping'') and are interpreted 
as dyons. The marked objects ``D'' and ``CAL'' in (b) are portrayed in 
detail in Fig.~\ref{fig:portraits}.}
\label{fig:event_pattern}
\end{figure*}

Thus, for each boundary condition, we will search for peaks of the modulus
of the corresponding fermionic topological charge density. In addition, in
order to define a size for the charge cloud surrounding the peaks, the respective
topological charge density is separately subjected to a cluster analysis. 
As usual (see Ref.~\cite{Ilgenfritz:2004zz,Ilgenfritz:2006ju,Ilgenfritz:2007xu}) 
the cluster analysis is a procedure to identify {\it connected} clusters among 
those lattice sites $x \in {\cal S}$, that have been selected by the condition 
that the modulus of the topological charge density $|q(x)|$ exceeds a certain 
threshold value $q_{\rm cut}$. Two sites $x,y \in {\cal S}$, being neighbors 
on the lattice, belong to the same cluster, if the signs of $q(x)$ and $q(y)$ 
agree. Otherwise they belong to different clusters. Guided by 
Ref.~\cite{Bruckmann:2006wf}, the threshold is chosen relative to the maximal 
density in the configuration as $q_{\rm cut} = \frac{1}{5} \max_x(|q(x)|)$,
safely above the point where the clusters coalesce and, finally, percolate. 

\begin{figure*}[!htb]
\begin{center}
\includegraphics[width=.4\textwidth]{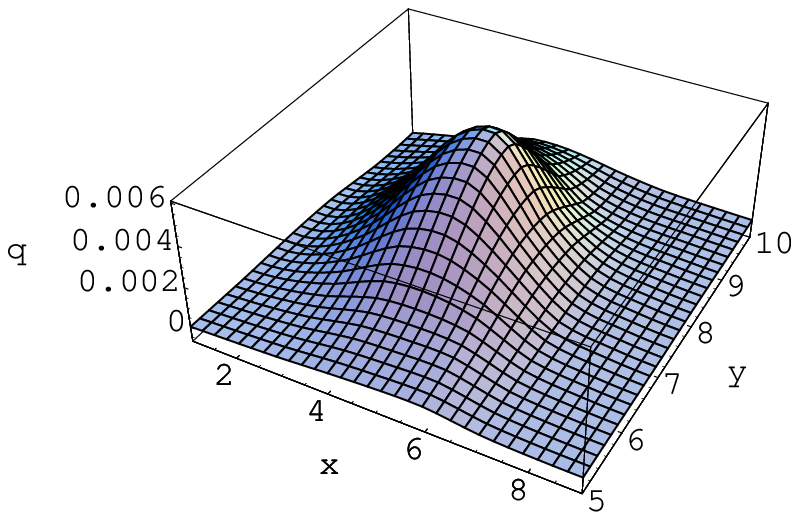}%
\hspace{1cm}\includegraphics[width=.4\textwidth]{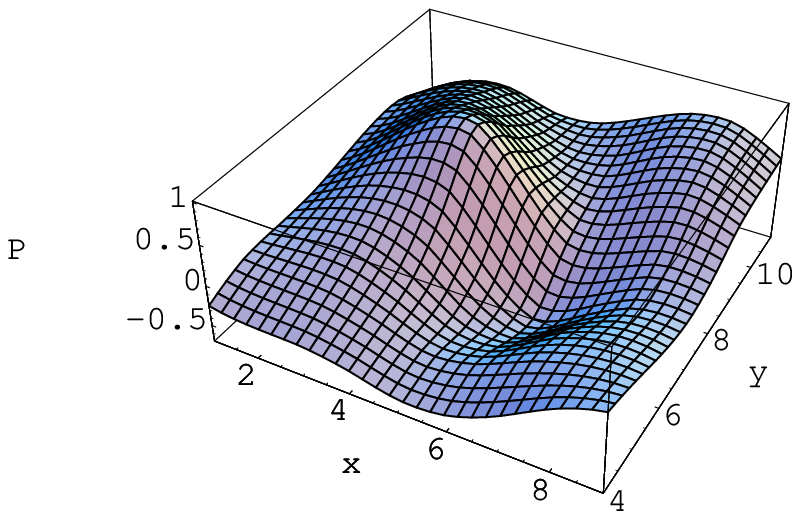}\\
\Large{(a)}\\
\vspace{1cm}
\includegraphics[width=.4\textwidth]{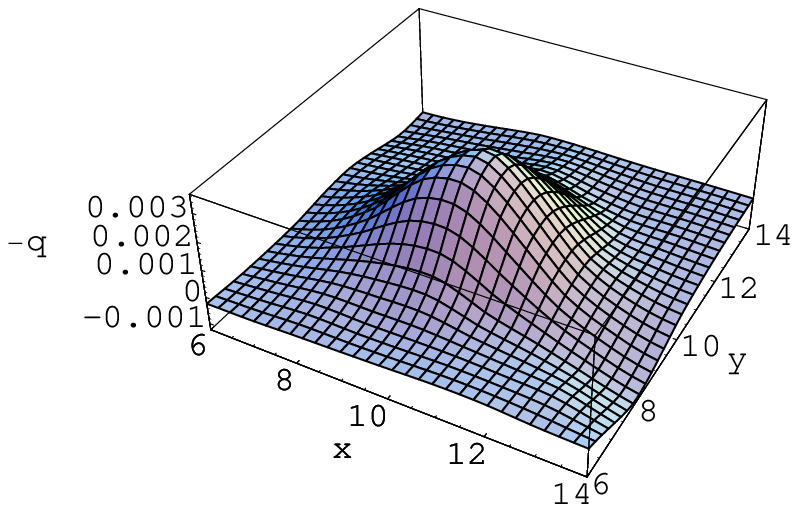}%
\hspace{1cm}\includegraphics[width=.4\textwidth]{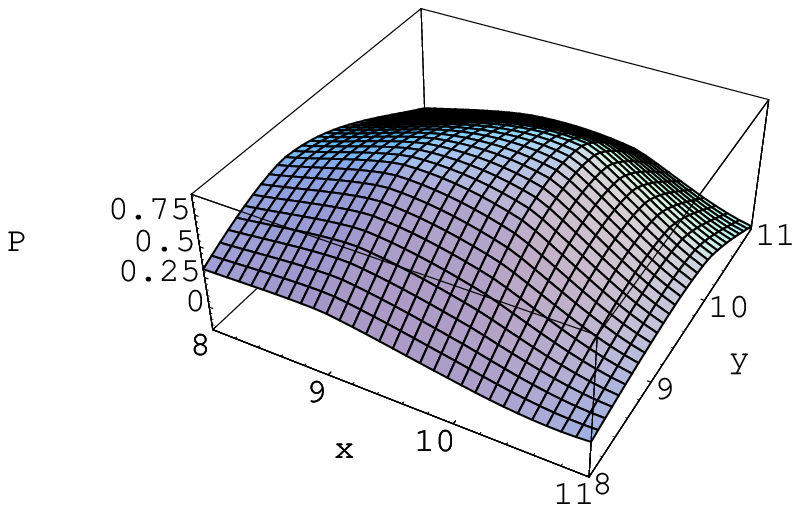}\\
\Large{(b)}\\
\end{center}
\caption {The fermionic topological charge density $q^{(p/a)}(x)$ (left) 
and the Polyakov loop $p({\vec x})$ (right): (a) for a typical caloron cluster 
(when $q^{(p)}(x) \approx q^{(a)}(x)$) and (b) for a typical dyon cluster
(which was visible only in $q^{(p)}(x)$) from Fig.~\ref{fig:event_pattern} (b). 
The topological density and the Polyakov loop are represented as function over 
part of the $(x,y)$-plane.  Please notice the different scales for the topological 
charge density and for the Polyakov loop. 
The Polyakov loop is measured after $N_{\rm APE}=10$ smearing steps.}
\label{fig:portraits}
\end{figure*}

For the set ${\cal C}^{(b)}$ ($b=p,a$) of clusters $c^{(b)}_i$ in a configuration 
found by the cluster analysis of the two densities $q_{\lambda_{\rm cut}}^{(p)}(x)$ 
and $q_{\lambda_{\rm cut}}^{(a)}(x)$, respectively~\footnote{For convenience we 
simplify the notation from now by dropping the subscript $\lambda_{\rm cut}$
from $q_{\lambda_{\rm cut}}^{(b)}$.}, 
we record the maximal value of the modulus of the corresponding density, 
\begin{equation}
|q^{(b)}_{{\rm max~cluster},i}| = \max_{x \in c^{(b)}_i} |q^{(b)}(x)| \; ,
\end{equation}
the sign ${\rm sgn}(q^{(b)}_{{\rm max~cluster~}i})$ and the corresponding 
space-time position $x$ of the peak inside each cluster $c^{(b)}_i$. 
The main purpose of defining the finite size clusters around the peaks
is to characterize the behavior of the Polyakov loop in the vicinity.
The Polyakov loop is always measured after $N_{\rm APE}=10$ smearing steps. 
Although the sign of the Polyakov loop at the cluster centers (topological
density peaks) is found to be dictated by the temporal periodicity/antiperiodicity 
imposed on the Dirac operator, the Polyakov loop is monitored {\it all over the 
cluster} to give auxiliary information. Its extremal values, 
$P^{(b)}_{{\rm max~}i}$ and $P^{(b)}_{{\rm min~}i}$, inside the clusters 
$c^{(b)}_i$ are recorded.  At least one of the two corresponds to the fermionic
boundary condition that defines the clusters, being positive for the periodic 
boundary condition and negative for the antiperiodic boundary condition. 

Figs.~\ref{fig:event_pattern} (a) and (b) show ``cluster plots'' representing 
two typical lattice configurations. A cluster $c^{(b)}_i$ of the topological 
charge density $q^{(b)}(x)$ is represented in the cluster plot by a filled circle 
($c^{(p)}_i$) for the periodic boundary condition or by a filled triangle 
($c^{(a)}_i$) for the antiperiodic boundary condition. The clusters 
$c^{(b)}_i$ are plotted in Figs.~\ref{fig:event_pattern} at the appropriate position 
\begin{equation}
(q^{(b)}_{{\rm max~cluster},i}~,~P^{(b)}_{{\rm extr~cluster},i}) 
\end{equation}
in the $(q_{\rm max~cluster},P_{\rm extr~cluster})$ plane. Here 
$P_{{\rm extr~cluster},i}$ denotes either $P^{(p)}_{{\rm max},i}$ or 
$P^{(a)}_{{\rm min},i}$, according to the $p$ or $a$ boundary condition that 
has defined the cluster through the corresponding topological charge density. 
Notice that this means that all circles appear in the upper and all triangles 
in the lower half-plane. 

Sometimes it happens that after changing the boundary condition from periodic 
to antiperiodic, one of the new clusters, $c^{(a)}_i$, nearly coincides in its 
space-time position $x$ with one of the previous ones, $c^{(p)}_j$, with a shift 
of the peak position less than a distance $d=2a \approx 0.22 {~\rm fm}$ in 
space-time. This would correspond to the ``not jumping'' case of 
Ref.~\cite{Gattringer:2002tg} where, however, only the scalar density of a single 
zero mode was under consideration. In this case, such clusters, the circle 
$c^{(p)}_j$ and the triangle $c^{(a)}_i$, are connected in 
Fig.~\ref{fig:event_pattern} by a broken line. The numbers close to the lines
denote the approximate shift (0 or 1 or 2) of the peak position. 
Such a pair represents a complete 
``caloron'', and the average over the respective topological charge densities 
$q^{(p)}(x)$ and $q^{(a)}(x)$ locally represents the true topological charge 
density inside the caloron. For calorons the topological charge clusters
defined for both types of boundary conditions are such that inside the clusters 
the extremal values of the Polyakov loop, $P^{(p)}_{{\rm max},i}$ and 
$P^{(a)}_{{\rm min},i}$, have clearly an opposite sign, indicative for the
dipole structure of a caloron in terms of the Polyakov loop.

Clusters that remained unpaired in this ``cluster plot'' have appeared only 
once, under only {\it one type} of boundary condition, such that the peak 
position could not be identified with a peak of the opposite boundary condition,
within a tolerance $d < 2a$. This corresponds to the ``jumping'' case of 
Ref.~\cite{Gattringer:2002tg}. Such clusters do not have an obvious partner 
(with opposite sign Polyakov loop and same sign topological charge density) 
suitable to form a ``caloron''. The length of the broken line attached to the
unpaired filled symbols represents the difference between the maximum and the 
minimum of the Polyakov loop inside the cluster. Numbers close to the unconnected 
lines denote the approximate distance (in the example, 3 or 4) between the cluster 
centers.
In contrast to the caloron clusters both maximum and minimum of the Polyakov loop
in an unpaired cluster are mostly of the same sign. In the few remaining cases the 
wrong-sign extremum is close to zero. Such clusters are 
called ``dyons'' because they, like the dyons in the classical $Q=2$ caloron 
solution shown in Fig.~\ref{fig:Q=2-caloron}, are invisible to the fermions 
under the ``wrong'' boundary condition.

The open circles around the filled symbols in the plots emphasize clusters 
which would have been localized knowing the zero mode(s) alone. These can 
also be clusters that we have to classify as calorons and as dyons. If they 
exist in the same configuration, these are clusters of a unique sign of the 
topological density, in accordance to the (yet unexplained) empirical fact 
that all zero modes of one configuration have the same sign of 
chirality.~\footnote{The cases of more than one zero mode per configuration
were excluded from the analysis in Ref.~\cite{Gattringer:2002tg}.}

Two characteristic objects that have been marked in
Fig.~\ref{fig:event_pattern} (b) as ``CAL'' and ``D'' are visualized in
Fig.~\ref{fig:portraits} in magnified form by their fermionic topological
charge density profile $q^{(p/a)}(x)$ (left) and their Polyakov loop
profile $p({\vec x})$ (right) within the occupied part of an $x$-$y$ section:
(a) for the caloron possessing the characteristic dipole structure of the Polyakov
loop and (b) for the dyon possessing a broad maximum of the Polyakov loop.
Let us stress that these objects have been identified in a generic Monte Carlo
lattice configuration without cooling or smearing. To be sure, the
Polyakov loop $p({\vec x})$ is presented after 10 smearing steps, which explains
the smooth picture.

\begin{figure*}[!htb]
\begin{center}
\includegraphics[width=.6\textwidth]{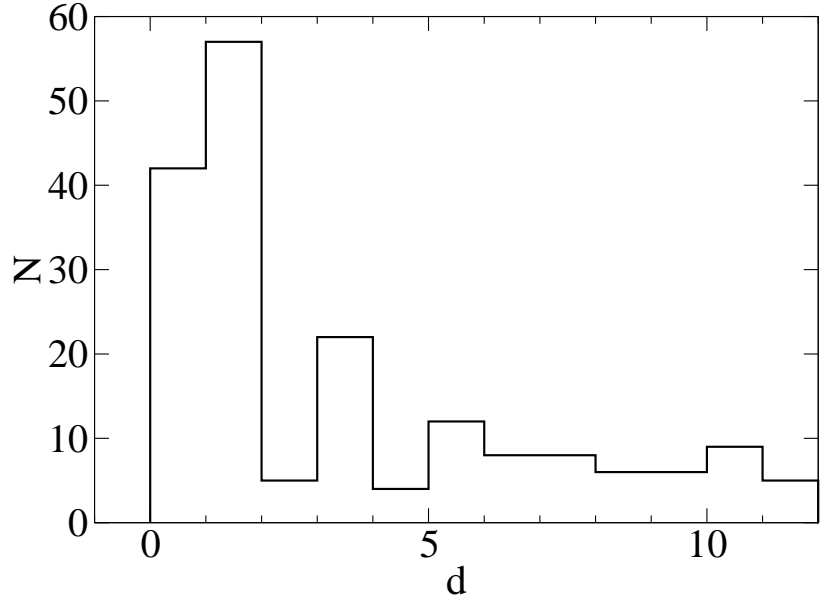}
\end{center}
\caption{The histogram of dyon-dyon distances in lattice units. 
The first two bins correspond to calorons which are unambiguously 
paired within distances $d < 2a$. The rest of the histogram with
$d \geq 2a$ refers to the remaining lumps grouped 
in suitable dyon-dyon pairs according to the closest distance.}
\label{fig:distance_histogram}
\end{figure*}
\vspace{1cm}

We have also studied the relative separation of appropriate dyon pairs. 
In Fig.~\ref{fig:distance_histogram} an histogram of dyon-dyon distances
in our sample is presented.  The first two bins correspond to calorons
with distances $d < 2a$. The rest of the histogram with $d \geq  2a$ contains 
pairs of suitably fitting dyon-dyon pairs, i.e. with the same sign of $q(x)$ 
and an opposite sign of $p({\vec x})$, grouped into pairs according to the 
closest distance. The statistics does not warrant so far the comparison
with a specific model for the caloron/dyon plasma.

\begin{figure*}[!htb]
\begin{center}
\includegraphics[width=.7\textwidth]{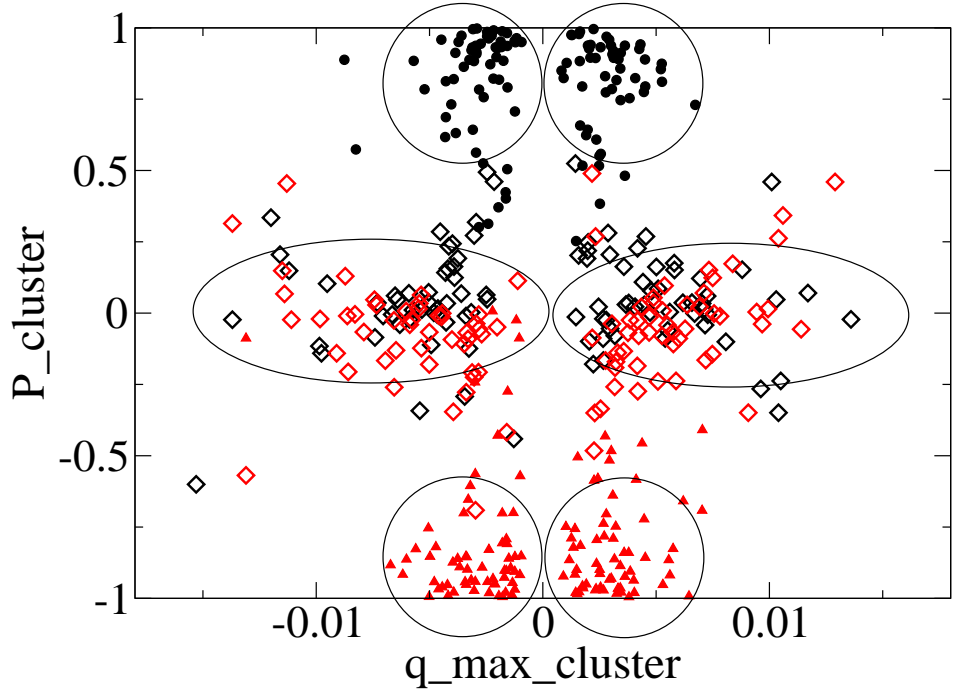}
\end{center}
\caption{The topological clusters of the whole sample shown analogously
to Fig.~\ref{fig:event_pattern}. 
The unpaired dyons are placed at their original 
$(q_{\rm max~cluster},P_{\rm extr~cluster})$ positions. 
The dyon pairs identified as calorons are finally re-located according 
to the average $\overline{P}_{\rm cluster}$ over their original, 
opposite sign values $P_{{\rm max}}$ and $P_{{\rm min}}$.} 
\label{fig:final_plot}
\end{figure*}

Finally, we have collected in Fig.~\ref{fig:final_plot} all clusters of the 
whole sample of 20 configurations analogously to Fig.~\ref{fig:event_pattern}. 
The unpaired dyons are placed at their original positions in the 
$(q_{\rm max~cluster},P_{\rm extr~cluster})$ plane. 
The difference to Fig.~\ref{fig:event_pattern} is that the two clusters 
close in space-time corresponding to an undissociated caloron in this plot 
are re-located according to the {\it average of the Polyakov loop} assigned 
to the respective {\it cluster} as follows, 
\begin{equation}
\overline{P}^{(b)}_{{\rm cluster},i}=
\frac{1}{2}\left(P^{(b)}_{{\rm max},i} + P^{(b)}_{{\rm min},i}\right) \; ,
\label{eq:averPL}
\end{equation}
a value which is scattered around zero because of the dipole structure.  
Thus, in Fig.~\ref{fig:final_plot}, each undissociated caloron is still 
represented by a close pair (with $b=p$ and $b=a$) 
of open squares, now with $|\overline{P}^{(b)}_{{\rm cluster},i}| < 0.25$.
The re-location according to the averaged Polyakov loop following
Eq. (\ref{eq:averPL}) leads in this scatter plot of clusters 
to a separation of points representing calorons and anticalorons (clustered 
in the ellipses) from the four types of dyons (clustered in the four circles)
with $|\overline{P}^{(b)}_{{\rm cluster},i}| > 0.5$.

The total number of isolated dyons (separated by a distance $d > 2a$)
is 113 plus 126 in this ensemble, whereas the number of dyons confined 
inside calorons (with a distance $d \leq 2a$) amounts to 101 plus 101,
meaning that on average approximately 10 calorons (dissociated or not)
are present per configuration,
if the resolution corresponds to 20 overlap eigenmodes. We emphasize
again that the counting is a counting of peaks. All peaks get classified
as dyons, regardless whether isolated or confined in calorons.
The total number corresponds to a caloron (or dyon pair) density 
$n^{1/4} = 230 {\rm~MeV}$. This in the right ballpark set by the topological
susceptibility, given the relative arbitrariness of the number of filtering 
modes. 

\section{Conclusion}
\label{sec:conclusion}

In this paper we have continued our search for specific KvB caloron-like
features in finite-$T$ lattice configurations. In a feasibility study
we have for the first time employed overlap valence fermions for this 
diagnostic purpose. 
More specifically, we have employed the dependence of eigenvectors and eigenvalues 
on the temporal boundary conditions imposed on the Dirac operator that can be 
changed at will. Thereby,
we have taken into consideration not only the zero mode(s) but the UV filtered 
topological charge density restricted to the 20 lowest modes per configuration.
The dependence of the apparent caloron/dyon content on the number of eigenmodes
has still to be systematically looked for.
According to Ref.~\cite{Bruckmann:2006wf} a resolution provided by 20 lowest 
fermionic eigenmodes, roughly corresponds to an amount of smoothing between 
10 and 20 smearing steps. 

In our previous work~\cite{Ilgenfritz:2004zz,Ilgenfritz:2006ju} we have
used smearing and the corresponding gluonic topological density. The amount
of smearing was, also somewhat arbitrarily, defined by the requirement that
the string tension should not drop below 60 \% of the full string 
tension~\cite{Ilgenfritz:2006ju} which allowed for 50 or 25 ... 20 smearing
steps in the confinement or deconfinement phase, respectively. 
Even more arbitrarily, the threshold for the definition of the clusters was 
set such that the density is split into a maximal number of clusters.
Under these circumstances, a large number of shallow clusters entered the
investigation before only a small part of the clusters could be successfully 
characterized -- by the monopole content -- as calorons or dyons. 

In this work, apart from the number of modes dictated by the PC memory, 
we have fixed the cutoff $q_{\rm cut}$ in a region where the number of 
clusters does not change with the cutoff and the size changes slowly. 
Moreover, the cluster centers were localized by the peaks of the modulus 
of the fermionic topological density $|q_{\lambda_{\rm cut}}(x)|$ and do 
not change anymore with the cutoff. Thus, the number of clusters is 
determined essentially by the number of analysing modes that was adopted 
in anticipation of a physically acceptable density of dyon pairs. What we 
could show here is that with this resolution the cluster composition of 
the topological charge can be understood in terms of calorons and dyons 
without serious problems.

All these clusters, once found,  are seen to be accompanied either by a dipole 
structure in the Polyakov loop $p({\vec x})$ or a broad maximum of the modulus 
of the Polyakov loop $|p({\vec x})|$. This shows that by means of the two 
topological densities (corresponding to periodic or antiperiodic temporal 
boundary conditions for overlap fermions) the task can be solved to identify 
calorons and dyonic constituents.  

In future investigations we will have to further specify those conditions for 
filtering that make the cluster charges distributed around $\pm 1$ and 
$\pm 1/2$ , hopefully a very stable result.
Furthermore, we hope for a better confirmation of the caloron/dyon model
by extending this study to lower temperature (where the model is good for
describing confinement) and to study also the higher temperature region.

\section*{Acknowledgements}
This work was partly supported by RFBR grants 05-02-16306, 06-02-04014 and 
06-02-16309 
and by the DFG grant 436 RUS 113/739/0-2 together with the RFBR-DFG grant 
06-02-04010.  
Three of us (V.G.~B, B.V.~M. and A.I.~V.) gratefully appreciate the support of 
Humboldt-University Berlin where this work was carried out to a large extent.
S.M.~M. is also supported by an INTAS YS fellowship 05-109-4821.
E.-M.~I. is supported by DFG (FOR 465 / Mu932/2).

\bibliographystyle{apsrev}
\bibliography{overlap_clusters.bib}

\end{document}